\documentclass[epj,nopacs]{svjour}
\usepackage{graphics}
\usepackage[OT1]{fontenc}
\usepackage[utf8]{inputenc}
\usepackage[english]{babel}
\usepackage{textcomp}
\usepackage{endnotes}
\usepackage{amstext}
\usepackage{amssymb}
\usepackage{amsmath}
\usepackage{graphicx}
\usepackage[unicode=true,pdfusetitle,bookmarks=true,bookmarksnumbered=false,bookmarksopen=false,breaklinks=false,pdfborder={0 0 1},backref=false,colorlinks=false]{hyperref}
\usepackage{breakurl}
\usepackage{cite}
\usepackage{wasysym}
\usepackage{array}
\usepackage[final,markup=nocolor]{changes}
\begin{document}
\title{Highly Confined Stacks of Graphene Oxide Sheets in Water}
\subtitle{}
\author{Rafael~Leite Rubim\inst{1} \and Margarida~Abrantes
  Barros\inst{1,2} \and Thomas~Missègue\inst{1} \and
  Kévin~Bougis\inst{1}\thanks{\emph{Present address:}
    Solvay--Laboratoire du futur, 178 avenue du Docteur-Schweitzer,
    F-33608 Pessac, France} \and Laurence~Navailles\inst{1} \and
  Frédéric~Nallet\inst{1} 
}                     
%
%
\institute{Universit\'e de Bordeaux, Centre de recherche
  Paul-Pascal--CNRS, 115 avenue du Docteur-Schweitzer, F-33600 Pessac,
  France \and T\'ecnico Lisboa, Av. Rovisco Pais, 1, PT-1049-001
  Lisboa, Portugal}
\mail{\texttt{frederic.nallet@u-bordeaux.fr}}
\date{Accepted for publication 2018-02-21}
\abstract{%
  Since the discovery of graphene oxide (GO), the most accessible of
  the precursors of graphene, this material has been widely studied
  for applications in science and technology. In this work, we
  describe a procedure to obtain GO dispersions in water at high
  concentrations, these highly dehydrated dispersions being in
  addition fully redispersible by dilution. With the availability of
  such concentrated samples, it was possible to investigate the
  structure of hydrated GO sheets in a previously unexplored range of
  concentrations, and to evidence a structural phase
  transition. Tentatively applying models designed for describing the
  small-angle scattering curve in the Smectic~A (or L$_{\alpha}$)
  phase of lyotropic systems, it was possible to extract elastic
  parameters characterising the system on the dilute side of the
  transition, thereby evidencing the relevance of both electrostatic
  and steric (\textsc{Helfrich}) interactions in stabilising aqueous
  lamellar stacks of GO sheets.
%
} 
\maketitle
\section{Introduction}
\label{intro}
Graphene oxide (GO) is a material obtained by mild oxidation and
exfoliation of graphite, and one of the most common manner of
preparing it is \textsc{Hummer}'s modified
method~\cite{hummers,synthesis}. This material is attracting a lot of
interest, in particular because it can easily be dispersed in various
solvents, including water, and many GO-based materials and composites
have been developed by solution
processing~\cite{Behabtu2010,Kim2011,Bisoyi2011,XuGao2011_2,Frost2012,Kim2012}.
\par The structure of GO sheets, as well as their structural
organisation in water have been investigated using various techniques,
including atomic force microscopy (AFM), polarised optical microscopy
(POM), circular dichroism (CD) and small-angle x-ray scattering
(SAXS)--see for
instance~\cite{Kim2011,XuGao2011_1,Aboutalebi2011,Dan2011,XuGao2011_2,Zamora-Ledezma2012,Cao2013,DiMauro2016}. Most
of these works--either directly (AFM) or indirectly (SAXS)--points to
an atomic thickness $t$ for the GO sheet, significantly below $1$~nm
(assuming for the GO density $\rho_{\mathrm{GO}}$ a value around
$1.8$~g/cm$^3$ in the SAXS-based method). Furthermore, it is now a
consensus from POM and SAXS studies that the phase diagram of the
lyotropic GO--water dispersion exhibits isotropic, nematic and
lamellar (or lamellar-like) phases, phase transitions being driven by
the increase of GO concentration in the dispersion, as also observed
in somehow similar materials made of planar, solid-like sheets of
near-atomic thickness~\cite{Gabriel2001,Michot2006}--or a bit
thicker~\cite{Kleshchanok2012}.
\par In qualitative accordance with \textsc{Onsager}'s theory for the
isotropic-to-nematic phase transition in suspensions of \emph{hard}
colloids, \added{quantitatively valid for slender
  particles}~\cite{Onsager1949}, the particle volume fraction
$\varphi_I$ at the transition onset is given \added{in numerical
  simulations for ``pancake'' particles
  by~}\cite{Frenkel1982,Cinacchi2014,Cinacchi2015}
\begin{equation}
  \label{eq:onsagerI}
  \varphi_I\approx3.2\frac{t}{L}
\end{equation}
in terms of the particle aspect ratio
$L/t$. Equation~(\ref{eq:onsagerI}) describes reasonably well the
\emph{mass} fraction $f_m^B$ of GO when birefringence first occurs
(\emph{viz.} when the nematic phase first appears) using
$f_m^B=\varphi_I\times\rho_{\mathrm{GO}}/\rho_{\mathrm{H}_2\mathrm{O}}$,
considering the dispersity \DJ\ in lateral extensions
$L$~\cite{Kooij1998,Bates1999}, as well as uncertainties in GO
thickness $t$ and density $\rho_{\mathrm{GO}}$. Such an agreement is
considered as a convincing argument for the GO sheets being
\emph{rigid} enough to remain essentially uncrumpled in dilute
suspensions~\cite{Kim2011,XuGao2011_2}.
\par \added{Structures commonly described as }\emph{lamellar} are
observed in more concentrated GO dispersions, as mainly results from
SAXS studies~\cite{XuGao2011_1,XuGao2011_2,Zamora-Ledezma2012,Xu2014},
a behaviour also found in similar (inorganic) materials such as
phosphatoantimonates, clays or \added{titanium-iron acid
  oxides~}\cite{Gabriel2001,Michot2006,Kleshchanok2012,Geng2013}. The
structure (sometimes depicted \added{more cautiously} as a nematic
gel, a locally layered system, or a pseudo-smectic phase
\added{because compelling evidence for \emph{positional} long-range
  order is not easily found}) is formed by stacking GO sheets (or
other kinds of solid-like sheets), separated by layers of water, with
a given distance of repetition $\ell$ of the unit cell along the
stacking axis $z$. In the plane perpendicular to $z$, the structure of
the two-dimensional solid-like sheet is well-defined locally, but more
difficult to ascertain at scales larger than $L$. Owing to the
repulsive interaction along $z$ between two facing sheets, with a
significant electrostatic contribution according to
refs.~\cite{Gabriel2001,Michot2006,XuGao2011_2}, the thickness of the
water layers increases, with therefore an increase in $\ell$, when
(low ionic strength) water is added to the system. In a geometric
description of the swelling process where $L\rightarrow\infty$ and $t$
is a constant, a simple dilution law, namely
\begin{equation}
\ell=t/\varphi
\label{eq:1Ddil}
\end{equation}
is expected and indeed observed, as in
refs.~\cite{XuGao2011_2,Zamora-Ledezma2012}, at least for a restricted
range of particle volume fractions $\varphi$,
see~\cite{Gabriel2001,Michot2006}. The dilution law,
eq.~(\ref{eq:1Ddil}), yields the above-mentioned SAXS (indirect)
estimate for the GO sheet thickness $t$.
\par One of the purpose of the present contribution is to explore the
validity of eq.~(\ref{eq:1Ddil}) towards more \emph{concentrated} GO
dispersions than previously investigated. In the next sections, we
describe how our samples are characterised using dynamic light
(sect.~\ref{sampleCharacterisation}) or small-angle x-ray
(sect.~\ref{techniques}) scattering techniques, and dehydrated in a
controlled way to almost complete dryness while remaining fully
redispersible in water (sect.~\ref{samplePreparation}). Our main
results are summarised in sect.~\ref{res}, with evidences for a
structural phase transition between lamellar structures, not reported
previously, as dehydration proceeds. In sect.~\ref{disc}, we discuss
possible mechanisms stabilising in water the lamellar stacks of GO
sheets, drawing an analogy with lamellar stacks of self-assembled
amphiphilic bilayers. Though we give no clues as regards the most
concentrated regime, the lamellar stack of GO sheets appears to be
well described in the dilute regime by the so-called ``unbinding
transition'' phenomenon that results from repulsive
\textsc{Helfrich}~\cite{Helfrich1978} and electrostatic interactions
between stacked layers competing with attractive \textsc{van der
  Waals} interactions~\cite{Lipowski1986,Milner1992}.
\section{Materials and methods}
\label{MatMet}
\subsection{Sample characterisation}
\label{sampleCharacterisation}
\begin{figure}[h]
\begin{center}
\includegraphics[width=0.45\textwidth]{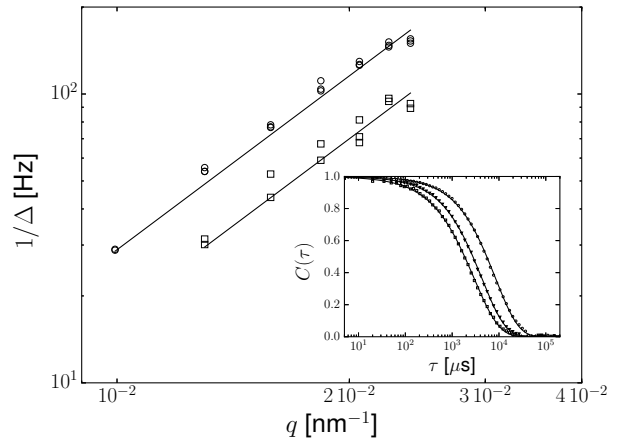}
\caption{Characteristic relaxation frequency $\Delta^{-1}$ of the
  autocorrelation function measured in DLS as a function of the
  scattering vector $q$ for GO dispersions prepared from two separate
  batches--batch~1: $\circ$, batch~2: $\square$. Inset:
  Stretched-exponential model fitted to selected DLS data.
\label{fig:dls}}
\end{center}
\end{figure}

The graphene oxide suspensions are prepared from a commercial aqueous
solution sold by Graphenea (San Sebastian, Spain), with nominal
concentration 4~mg/mL. \added{Such a solution is concentrated enough
  to be birefringent, as revealed by POM, but does not yet exhibit any
  significant increase in viscosity compared to water.} According to
the producer, the dispersion presents more than 95~\% of carbon
monolayers and an amount between 41 and 50\% of oxygen atoms, with
variable sheet dimensions $L$ below 10~$\mu$m, usually around
1--2~$\mu$m~\cite{graphenea}. \added{Owing to the presence of COOH
  groups attached to the sheet surface, the aqueous GO suspensions are
  expected to be acid and, indeed, their measured $p$H is about 2.4.}
Two different batches were bought and used to prepare the samples. For
both batches, dynamic light scattering (DLS) experiments were carried
out, using a research goniometer and laser light scattering system
from Brookhaven Instruments Corporation (Holtsville, NY, USA). Freshly
prepared samples were diluted in water to 0.04~mg/mL. \added{At such a
  concentration, the samples are no longer birefringent but faint
  depolarised fluctuations can be observed by POM in an optically
  thick ($0.800$~mm) cell from VitroCom (Mountain Lakes, NJ, USA). The
  DLS experiment is performed with incident light polarised
  per\-pen\-di\-cu\-lar to the scattering plane, without analysing the
  polarisation of the scattered signal.} DLS experiments have been
repeated from time to time on ageing samples prepared with the first
batch along a total period of about 2 months in an attempt to
characterise ageing, if any. Some representative results on freshly
prepared samples are shown in fig.~\ref{fig:dls}.
Fitting the DLS data to a stretched exponential model, see
eq.~\ref{eq:correlation}, as a convenient (but \emph{ad hoc}) way
to somehow take into account the GO dispersity, two parameters (a
characteristic time and a stretching exponent) were obtained as a
function of the scattering wave vector $q$.
\par The model correlation function is expressed as
\begin{equation}
  C\left( \tau \right) = \exp \left[ -2 \left( \frac{\tau}{\Delta}
    \right) ^{\beta} \right]
\label{eq:correlation}
\end{equation}  
where $\Delta$ is the characteristic relaxation time and $\beta$ the
stretching exponent. Parameter $\beta$ was found to decrease from
\emph{ca.} 0.9 to 0.7 as the scattering vector $q$ increases from
\emph{ca.} 1$\times10^{-2}$ to 2.2$\times10^{-2}$~nm$^{-1}$. Besides,
as illustrated in fig.~\ref{fig:dls} by the straight lines with a
slope 2, the relaxation frequency $\Delta^{-1}$ is proportional to
$q^2$, meaning that an \emph{effective} diffusion coefficient--or a
hydrodynamic radius $R_H$--can be defined. From the standard
Stokes-Einstein relation
\begin{equation}
  \label{eq:stokes-einstein}
  R_H=\frac{k_BT\Delta}{6\pi\eta}\times q^2
\end{equation}
with $k_B$ the Boltzmann constant, $T$ the absolute temperature of the
GO dispersion and $\eta$ the solvent viscosity, hydrodynamic radii
were found equal to 0.74~$\mu$m and 1.22~$\mu$m for batches~1 and 2,
respectively.
\par As mentioned above, possible effects of ageing were checked on
samples prepared from the first batch, with three distinct histories:
\begin{enumerate}
\item Samples were diluted to the concentration appropriate for DLS
  (\emph{ca.} 0.04~mg/mL) \emph{immediately} after receiving the
  solution from Graphenea, then stored for ageing;
\item Samples were diluted to the concentration appropriate for DLS
  from the solution \emph{stored for ageing} received from the
  manufacturer;
\item Samples \emph{concentrated} to \added{$\approx160$~mg/mL}
  immediately after reception of the Graphenea solution (see below,
  sect.~\ref{samplePreparation}, for details regarding the
  concentration procedure) were stored for ageing, then diluted to the
  concentration appropriate for DLS.
\end{enumerate}
Whatever the sample history, the storage conditions were the same,
namely stable temperature (22$^{\circ}$C) and no exposure to direct
light. In all cases, DLS did not reveal any significant ageing over a
period of about 2~months.
\subsection{Sample preparation}
\label{samplePreparation}
\begin{figure}[h]
\includegraphics[width=0.45\textwidth]{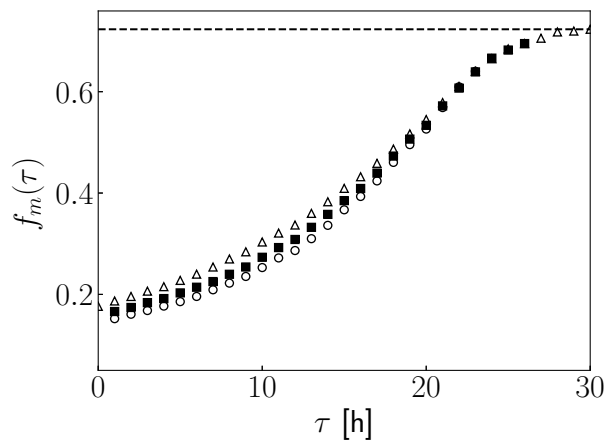}
\caption{GO mass fractions $f_m(\tau)$ as a function of the
  dehydration time in the vacuum chamber ($p=300$~mbar) for three
  samples differing by their \emph{initial} water content:
  \added{$\circ$ 14.4~\%, $\blacksquare$ 15.7~\%, $\vartriangle$
    17.6~\%. The horizontal dashed line at $f_m=0.723$ corresponds to
    the limiting GO mass fraction, and accounts for water molecules
    that cannot be removed with our drying set-up.}
  \label{fig:dehydration}}
\end{figure}

A procedure to increase the concentration of the commercial GO
dispersions was implemented, requiring two steps. Centrifugation and
ultracentrifugation are used in the first step. The commercial
dispersion is first centrifuged for about 20~min at 1400$g$, in order
to remove ``large'' aggregates from the sample. After discarding the
bottom phase, the supernatant is then ultracentrifuged at a much
higher speed (302000$g$) for 5~h. The recovered supernatant, mostly
water \added{at $p$H=2.6}, occupying almost the total volume of the
centrifuge cell, is also discarded. The remaining phase appears as a
highly viscous material with a dark, almost black colour. As explained
below--see also fig.~\ref{fig:dehydration}--, it turns out that the GO
mass fraction $f_m$ achieved at this stage is around \added{0.16},
thus corresponding to an increase in GO concentration by a factor
about \added{40}. We have checked that increasing the duration of the
ultracentifugation procedure does not significantly increase $f_m$,
while decreasing it below \emph{ca.} 3~h \added{does not lead to
  concentrated enough dispersions}.
\par After a period of about one week left for homogenisation in a
closed container--required because the presence of small and
uncontrolled amounts of water at the surface of the concentrated
dispersion cannot be avoided when recovering the pellet from the
centrifuge cell--the second step begins. A home-designed device is set
up to \emph{slowly} evaporate at room temperature the aqueous solvent
from samples. It consists in a diaphragm pump connected to a
desiccator where a dozen of (open) Eppendorfs containing the desired
material are stored, with a pressure control system maintaining
300~mbar inside the vacuum chamber.
\par Figure~\ref{fig:dehydration} shows the evolution in time of the
GO mass fraction for three dispersions resulting from the first,
cen\-tri\-fugation-based step. As it appeared retrospectively, they
differed by their \emph{initial} mass fractions. The mass fraction
$f_m(\tau)$ ($\tau=0$ when dehydration begins in the vacuum chamber)
is determined, indeed, by weighing the sample at time $\tau$, which
obviously requires \emph{opening} the vacuum chamber. The measured
mass is $m(\tau)$. The clock is stopped (and the Eppendorfs closed)
for the duration $\delta$ of the weighing operations, with an optional
(mild) shaking intended to re-homogenise samples visually displaying
drier patches. The same procedure is repeated at regular intervals of,
typically, 1~h (in ``vacuum times'', \emph{i.e.}  subtracting the
$\delta$'s from the actually elapsed time). It has been observed that
for $\tau\gtrsim30$~h the mass $m(\tau)$ does not decrease any more,
and keeps a constant value $m_{\infty}$. We have checked on a few
sacrificial samples, submitted to a somehow stronger vacuum
\added{(pressure in the mbar range)} for about 15~h, that remaining
water cannot be extracted with our set-up: \added{Achieving complete
  dehydration would require ultra-vacuum or elevated
  temperatures}~\cite{Scholz1969,Lerf2006,Rouziere2017}. \added{On the
  basis of our x-ray measurements (see below, Section~\ref{res}), we
  estimate the weight fraction of ``bound'' water from
  Ref.~}\cite{Scholz1969} \added{to be $f_w^{\infty}\approx27.7$\%,}
with therefore $f_m(\tau)=(1-f_w^{\infty})\times m_{\infty}/m(\tau)$.
\par Notably--and similarly to the first step--, the second step of
the dehydration procedure is \emph{reversible}. As shown by SAXS (see
below, sect.~\ref{expRes_disc}), adding water to a sample extracted at
time $\tau_>$ from the desiccator in the required amounts to mimic the
composition of a sample stored for a lesser time $\tau_<$ leads to
essentially identical diffractograms for the ``wet'' and
``dried-rehydrated'' samples when they both originate from the same
centrifuged material. At contrast with what has been observed with,
\emph{e.g.}, freeze-dried GO dispersions where the GO chemical
structure is strongly affected~\cite{Ham2014}, it seems to be
preserved in our case as no aggregates were found in the slowly dried
samples redispersed in water.
\subsection{Experimental techniques}
\label{techniques}
Samples removed at time $\tau$ from the desiccator were left for at
least a week in their (now closed) preparation Eppendorfs to ensure
relaxation of possible humidity gradients. After homogenisation, the
samples were analysed by POM. Due to their extreme opacity when $f_m$
exceeds 50\%, highly dehydrated samples could not be successfully
observed. For samples with smaller mass fractions, images were
recorded (data not shown) using an Olympus BX 51 microscope with
crossed polarisers and a $\times20$ objective. The samples were
sandwiched between a glass slide and a cover-slip, without special
precautions for ensuring a constant optical path, \added{estimated
  below 10~$\mu$m}, but preventing water evaporation by means of a
UV-curing glue. Birefringence was always observed, indicating a
liquid-crystalline organisation. The samples were homogeneous, as
revealed by observing them without the analyser, indicating that
aggregates were not present.
The samples were also investigated by small angle x-ray
scattering. The thick pastes were spread on a circular (diameter
1.3~mm), machine-drilled opening \added{perpendicular to the long axis
  of} cylindrical stainless steel supports (2.0$\times$20.0~mm) which
were then introduced in quartz capillaries with a nominal diameter of
2.5~mm. The spreading procedure did not allow a control of the optical
path better than $\approx25$~\%. The quartz capillaries were further
flame-sealed, to ensure tightness. Diffractograms were recorded on a
Bruker-AXS Nanostar machine equipped with a Hi-Star detector, also
from Bruker (Karlsruhe, Germany). From the entrance pinhole to the
beryllium window in front of the detector, the whole flight path is
evacuated. A crossed-coupled pair of G\"obel mirrors (Bruker) selects
the $\lambda=1.5418$~\AA\ radiation of a copper source (Siemens),
operated at 40~kV and 35~mA. A 3-pinhole system is used for
collimating the incident beam, with a size (FWHM) at sample position
\emph{ca.}  0.43~mm in both vertical and horizontal directions. Two
sample-to-detector distances, found close to 0.25~m and 1.05~m
respectively, calibrated using silver behenate as
standard~\cite{Huang1993}, were used to match the variable stacking
periods of the samples. From the Gaussian width of the first order
Bragg peak of silver behenate, we estimate a resolution width (FWHM)
$\Delta q\approx5\times10^{-2}$~\AA$^{-1}$ or $\Delta
q\approx6.0\times10^{-3}$~\AA$^{-1}$ for the two configurations,
respectively. \added{Owing to the intrinsic broadening of silver
  behenate~}\cite{Huang1993}\added{, the latter value could be
  slightly over-estimated}. The scattering wave vectors that are
practically accessible after subtracting the signal of a reference
(water) capillary range from 0.04~\AA$^{-1}$ to 0.8~\AA$^{-1}$ in the
``large-angle'' configuration, and from 0.01~\AA$^{-1}$ to
0.2~\AA$^{-1}$ in the ``small-angle'' one. For accessing to even
higher scattering wave vector values (typically 0.5--3.3~\AA$^{-1}$),
as required to assess the \emph{in-plane} order of the GO sheets, we
use a custom-made instrument with a copper rotating-anode-based setup
and crossed-coupled pair of G\"obel mirrors, both from Rigaku (Tokyo,
Japan), a 3-pinhole collimation system similar to the Bruker one and a
mar345 image plate detector (marXperts, Norderstedt, Germany) with
sample-to-detector distance 0.15~m. At contrast with the Bruker
system, only the collimation flight path is evacuated. Acquisition
times on the instruments were in the order of 5 hours (Bruker
Nanostar) or 1 hour (custom instrument). Temperature, fixed at
20$^{\circ}$C, is controlled to within $\pm0.2^{\circ}$C by a water
circulation system (Bruker Nanostar) or, with a lesser precision, by
the air-conditioning system of the room (custom instrument). For both
instruments, the 2D detector images were most often characteristic of
slightly oriented samples, as previously
observed~\cite{XuGao2011_2,Zamora-Ledezma2012}, presumably because of
the shear applied when filling the x-ray capillaries, or spreading the
thick samples on the circular opening of the sample holders. Data was
therefore azimuthally averaged to yield (normalised) intensities $I$
\emph{vs.}  scattering wave vector $q$ curves.
\section{Experimental results and discussion}
\label{expRes_disc}
\subsection{Results}
\label{res}
\begin{figure}[h]
\begin{center}
\includegraphics[width=0.45\textwidth]{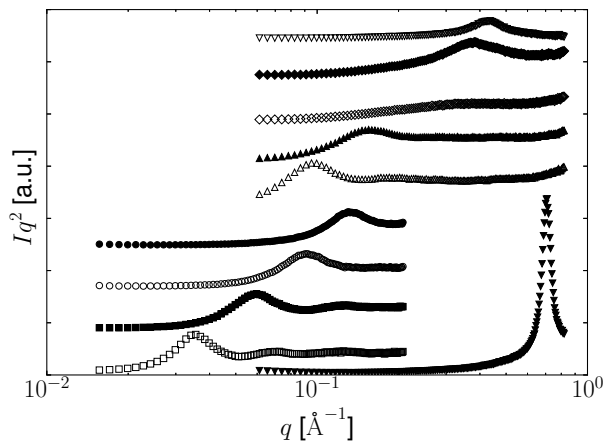}
\caption{SAXS spectra (Kratky plot: $Iq^2$ \emph{vs.} $q$) for GO
  aqueous dispersions differing by their GO mass fractions $f_m$:
  \added{0.04 ($\square$), 0.07 ($\blacksquare$), 0.10 ($\circ$),
    0.14 
    ($\bullet$), 0.12 ($\vartriangle$), 0.16 ($\blacktriangle$), 0.21
    ($\lozenge$), 0.26 ($\blacklozenge$), 0.38 ($\triangledown$) and
    0.62 ($\blacktriangledown$)}.  Data shifted vertically by amounts
  allowing a better visualisation
  \label{fig:RD20-30}}
\end{center}
\end{figure}

\begin{figure}[h]
\begin{center}
\includegraphics[width=0.45\textwidth]{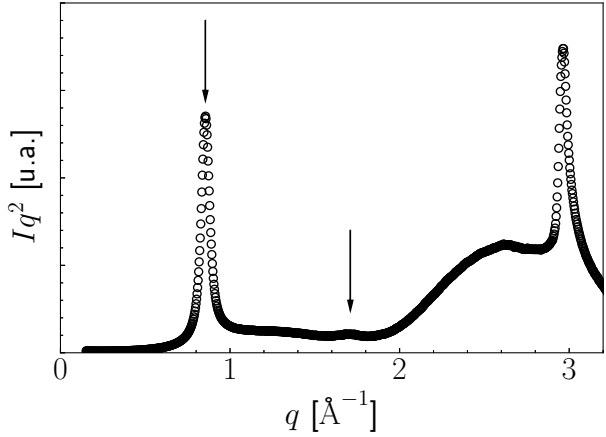}
\caption{X-ray scattering data for the dryiest GO dispersion
  \added{($f_m=0.792$). Lamellar stacking peaks marked by vertical
    arrows at $q_0=8.548\times10^{-1}$~\AA$^{-1}$ and
    $\approx1.71$~\AA$^{-1}$}. The 2D, in-plane order of the carbon
  atoms in GO sheets gives rise to the intense and thin peak observed
  at $q_{\mathrm{G}}=2.964$~\AA$^{-1}$. \added{Other intensity humps at
  $\approx2.6$~\AA$^{-1}$: Unidentified features, possibly related to
  experimental artefacts arising from background scattering}
  \label{fig:RR59_anode}}
\end{center}
\end{figure}

\begin{figure}[htb]
  \centering
  \includegraphics[width=0.45\textwidth]{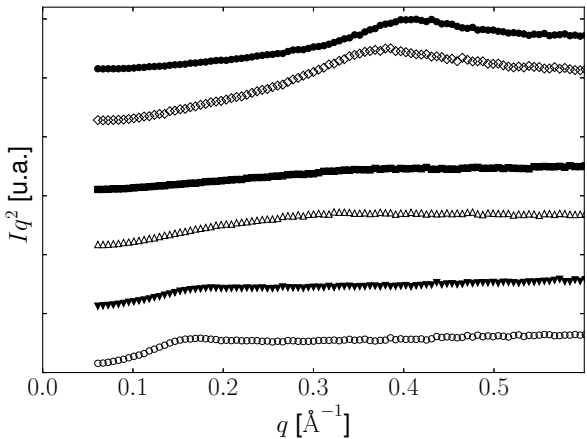}
  \caption{SAXS spectra in the Kratky representation for hydration
    values of the GO aqueous dispersions corresponding to a very broad
    first order Bragg peak. GO mass fraction $f_m$: \added{0.17
      ($\circ$), 0.19 ($\blacktriangledown$), 0.21 ($\triangle$), 0.23
      ($\blacksquare$), 0.26 ($\lozenge$), and 0.29 ($\bullet$)}.
    Data shifted vertically by amounts allowing a better
    visualisation}
  \label{fig:porodTransition}
\end{figure}

SAXS results (``small'' and ``large'' angle configurations) for ten
selected samples are shown for illustration in
fig.~\ref{fig:RD20-30}, in the $Iq^2$ Kratky representation that
factorises out the characteristic $1/q^2$ intensity decrease of very
extended, thin and flat particles with random
orientations~\cite{bookKratkyPorod1949}.
The observed peak, characteristic of the lamellar stacking, moves
towards higher scattering wave vector as dehydration proceeds. The
second order peak, though clearly observed in either the ``small'' or
``large'' angle configurations for the two more hydrated samples in
the corresponding series (GO mass fractions $f_m$ \added{0.04 and
  0.07, or 0.12 and 0.16}, respectively), barely appears in the
``large'' angle configuration for the other samples--even though it
still falls within the observation window. Nevertheless, as shown in
fig.~\ref{fig:RR59_anode} with an observation window extending to much
larger scattering wave vector values, the second order Bragg peak of
the lamellar stacking, though weak, is clearly observed in \added{one
  of the most dehydrated sample ($f_m=0.792$)}.
As a matter of fact, upon increasing the GO content up to
\added{$f_m\approx0.23$}, the intensity ratio between the second and
first order peaks decreases until the second order peak apparently
disappears, to be unambiguously recovered when $f_m$ reaches
\emph{ca.} \added{0.29}. In this concentration range, the first order
peak is also significantly \emph{broadened}--see
fig.~\ref{fig:porodTransition}.
Such features of the SAXS diffractograms may point to a structural
phase transition.  It is, however, not evidenced in the POM
observations. We return to this intriguing point immediately below.
\par
\begin{figure}[htb]
\begin{center}
\includegraphics[width=0.45\textwidth]{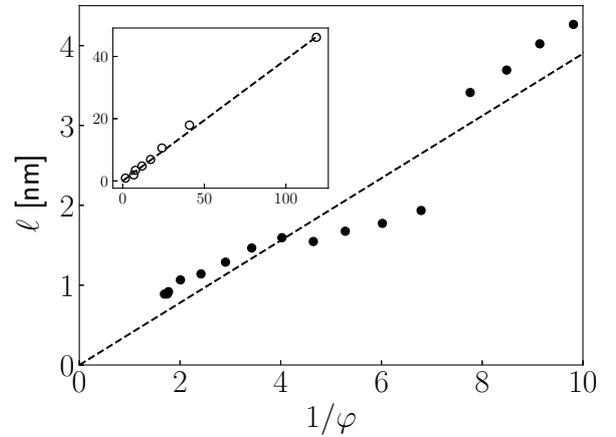}
\caption{Stacking period $\ell$ as a function of the inverse volume
  fraction $1/\varphi$ for highly dehydrated GO dispersions
  (\added{$0.1\leq\varphi$, or $0.17\leq f_m$)}. Inset: \emph{idem}
  for the whole dilution range. Dashed line: Simple swelling law
  $\ell=t/\varphi$, drawn with
  \added{$t=0.39$~nm}\label{fig:DilutionLine}}
\end{center}
\end{figure}

 From Bragg's law, namely $\ell=2\pi/q_0$, it is found that, as
expected, the period of the lamellar stack decreases when water is
removed from the structure. The experimental dilution law
$\ell(\varphi)$, with volume fractions $\varphi$ derived from mass
fractions $f_m$ through the relation
\begin{equation}
  \label{eq:phi-f}
  \varphi=\frac{\rho_{\mathrm{H}_2\mathrm{O}}f_m}{\rho_{\mathrm{H}_2\mathrm{O}}f_m+\rho_{\mathrm{GO}}(1-f_m)}
\end{equation}
(assuming volume additivity) is shown in fig.~\ref{fig:DilutionLine}.
A striking \emph{discontinuous} behaviour near
\added{$\varphi\approx0.14$ ($f_m$ close to 0.23)} is clearly observed
in the dehydrated limit of the dilution line. \added{Besides, the
  discontinuity is precisely found to occur in the hydration range
  where broadening of the first order peak, as well as the
  disappearance of the second order peak have been observed, hinting
  again at the occurrence of a structural phase transition. Still, }as
evidenced in the inset to fig.~\ref{fig:DilutionLine} where SAXS data
from samples submitted to the first concentration step
(sect.~\ref{samplePreparation})--some of them re-diluted--or only
mildly dehydrated in the second step has been included, our data
\added{remains broadly} compatible with the simple swelling law,
eq.~(\ref{eq:1Ddil}). This latter observation is nicely in agreement
with the findings of previous studies, limited then to significantly
more \emph{dilute} GO
dispersions~\cite{XuGao2011_2,Zamora-Ledezma2012} than investigated
here. The fit to the dilution data leads to a sheet thickness
\added{$t\approx0.39$~nm, a value close to, yet slightly lower than}
the value found in ref.~\cite{Zamora-Ledezma2012}. The structural
phase transition, if any, does therefore not strongly weakens the
relevance of the simple geometric arguments at the origin of the
simple swelling law.
\par As shown for illustration in fig.~\ref{fig:RR59_anode}
corresponding to \added{our dryiest sample ($f_m=0.792$)}, the
expected locally planar hexagonal structure of the carbon atoms in the
graphene layers is observed using x-ray scattering at large
angles. The peak found for \added{$q_G \approx$ 2.964~\AA$^{-1}$} can
be related to the C--C nearest-neighbour distance $d_{\mathrm{C-C}}$
in a given graphene layer using $3d_{\mathrm{C-C}}=4\pi/q_G$, which
indeed yields a result \added{(1.41~\AA) close to} the commonly
accepted value
$d_{\mathrm{C-C}}=1.42$~\AA\cite{StructureGraphite}. The same result
is found for more hydrated samples, as long as there is enough signal
for this peak to emerge from the background. 
%
\par \added{Two peaks (locations $q_0=0.85$ and 1.71~\AA$^{-1}$, close
  to $2q_0$) can be found in the lower $q$-range part of
  fig.~\ref{fig:RR59_anode}}. They are related to the lamellar
stacking order of the GO sheets. \added{The corresponding periodicity,
  about 0.74~nm, is \emph{ca.} $1.9\times$ higher than the geometric
  parameter $t\approx0.39$~nm found in fitting the simple dilution law
  to the whole set of SAXS data, that is to say about twice higher
  than the interlayer distance in
  \emph{graphite}~}\cite{StructureGraphite}. \added{This result is to
  be attributed to the water molecules remaining trapped between the
  GO sheets, about 46\% in volume fraction from eq.~(\ref{eq:1Ddil}),
  a value found in rather reasonable agreement with
  eq.~(\ref{eq:phi-f}), leading to $\varphi_w\equiv1-\varphi=0.32$.}
\subsection{Discussion}
\label{disc}
Despite a possible structural transition, occurring near
\added{$\varphi=0.14$} and remaining to be characterised in details,
it appears that GO aqueous dispersions exhibit a lamellar order over a
quite extended concentration range, with a stacking period $\ell$
varying from about 0.8~nm in the dryiest available system to more than
45~nm in our most hydrated samples. It is worth noting here that
periods as large as $\ell\approx100$~nm have even been found in other
studies~\cite{XuGao2011_2}. The physical mechanism stabilising the
lamellar structure for vastly different water contents is therefore of
obvious interest.
\par In the so-called lyotropic lamellar phases (self-assembled
bilayers of surfactant or lipid molecules separated by layers of
solvent, or solvent-swollen block-copolymer systems), a similar
swelling of the lamellar structure over very extended composition
ranges is also commonly
observed~\cite{Larche1986,Safinya1986,Strey1990,Freyssingeas1997,Ito2015,Uchida2016}. It
is similarly present in systems structurally similar to GO,
\emph{viz.}  based on extended solid-like
sheets--phos\-phato\-anti\-monate, for instance~\cite{Gabriel2001}, or
clay-based systems~\cite{Michot2006}--dispersed in aqueous
solutions. Such a swelling is commonly attributed to long-range,
either direct or effective, repulsive interactions between the stacked
sheets, acting across the solvent layers and of electrostatic origin,
or resulting from the ``undulation interaction'' mechanism proposed by
\textsc{Helfrich}~\cite{Helfrich1978}.
\par In the case of GO aqueous dispersions, the two mechanisms have
already been identified~\cite{XuGao2011_2,Poulin2016}, at least
indirectly in the case of \textsc{Helfrich}'s
mechanism~\cite{Poulin2016}. An electrostatic contribution is clearly
evidenced when experimentally studying the swelling properties in the
presence of added salts (in order to vary the ionic strength of the
aqueous solvent layers). Using NaCl as a typical univalent salt in the
\added{(nominal)} concentration range $10^{-6}$--$10^{-1}$~M, the same
effect as described in ref.~\cite{XuGao2011_2} is observed here,
namely a \emph{decreasing} stacking period $\ell$ with increasing salt
content above a $f_m$-dependent salt concentration $c_s^*$.
\begin{figure}[htb]
  \centering
  \includegraphics[width=0.45\textwidth]{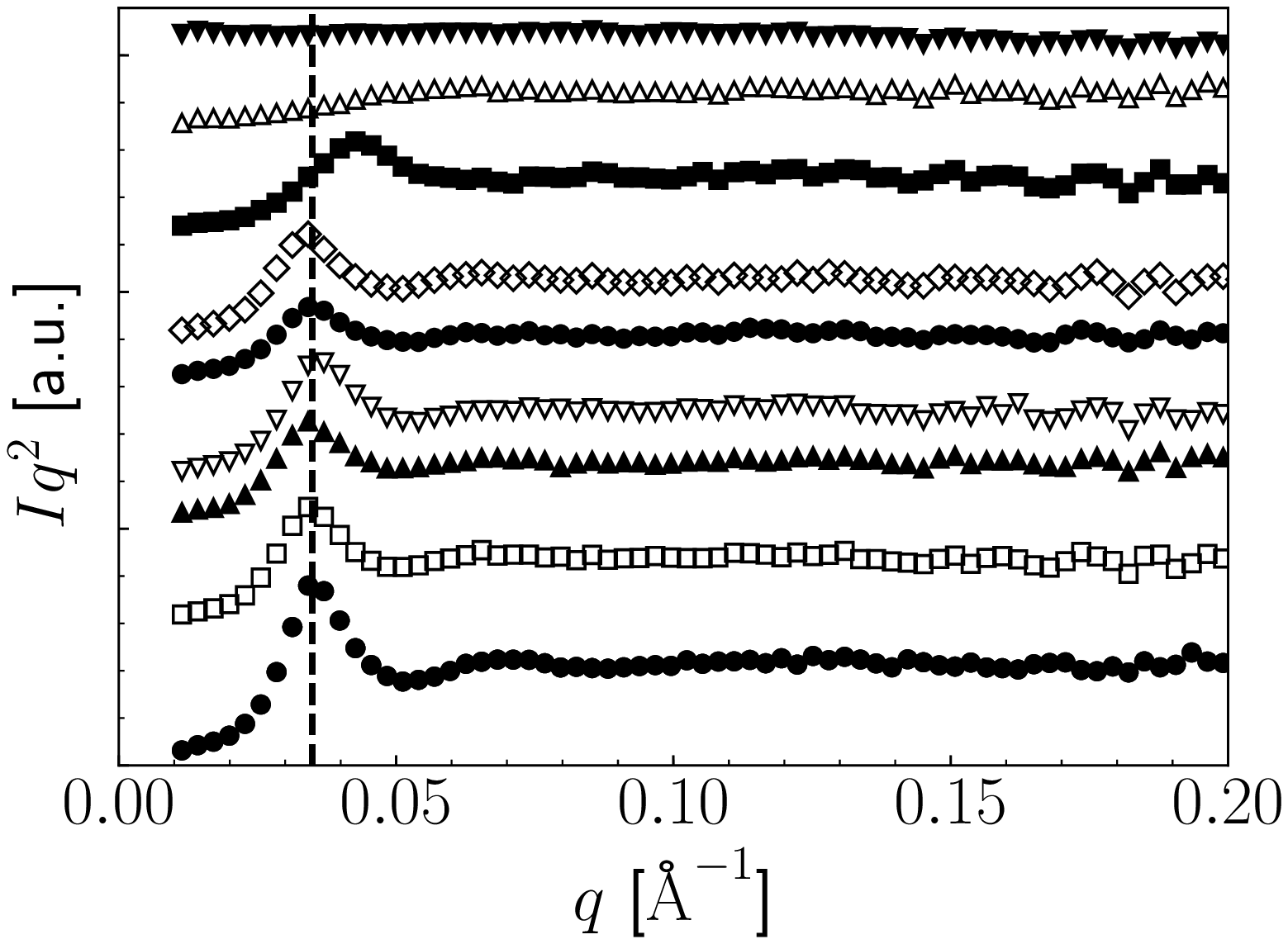}
  \caption{SAXS spectra in the Kratky representation for GO aqueous
    dispersions differing in added NaCl content $c_s$. GO mass
    fraction fixed to \added{$f_m=4.3\%$}. \added{Nominal} salt
    concentrations $c_s=1\times10^{-6}$~M ($\bullet$),
    $1\times10^{-5}$~M ($\square$), $5\times10^{-5}$~M
    ($\blacktriangle$), $1\times10^{-4}$~M ($\triangledown$),
    $5\times10^{-4}$~M ($\bullet$), $1\times10^{-3}$~M ($\lozenge$),
    $5\times10^{-3}$~M ($\blacksquare$), $5\times10^{-2}$~M
    ($\vartriangle$) and $1\times10^{-1}$~M ($\blacktriangledown$).
    Data shifted vertically by amounts allowing a better
    visualisation. The vertical dashed line is drawn at
    $q_0=3.49\times10^{-2}$~\AA$^{-1}$}
  \label{fig:NaCl}
\end{figure}
Indeed, as shown in fig.~\ref{fig:NaCl}, up to the salt concentration
$c_s^*$ (found about $1\times10^{-3}$~M for \added{$f_m=4.3\%$)}, the
first order Bragg peak position $q_0$ does not significantly change,
and the overall appearance of the SAXS spectra remains the
same. \added{Conversely,} the stacking period decreases and,
simultaneously, the first order Bragg peak broadens, then becomes
barely noticeable as the salt concentration increases above
$c_s^*$. Repeating the experiment at a different value for the GO mass
fraction \added{($f_m=1.4\%$)} yields qualitatively similar
observations (data not shown), with however a significant
\emph{decrease} in the value (about $1\times10^{-4}$~M) for $c_s^*$.
\par \added{Regarding now ``undulation interactions'', results have
  been interpreted in recent rheoSAXS experiments} by introducing a
bending modulus $\kappa$ for GO layers in the order of $k_BT$, that is
to say ``superflexible'' sheets~\cite{Poulin2016}. Such a low value
suggests quite strong steric repulsions between adjacent GO layers,
owing to the confinement of undulation fluctuations. This would nicely
explain the conspicuous swelling properties of the system and, in
particular, the salt effect mentioned above. Indeed, as has been
firmly established since \textsc{Helfrich}'s seminal
article~\cite{Helfrich1978}, swelling properties in lamellar stacks of
\emph{flexible} sheets result from a competition between, on one hand,
the ``unbinding'' tendency of undulation fluctuations and, on the
other hand, direct sheet--sheet interactions that may favour ``bound''
systems if attractive enough~\cite{Lipowski1986,Milner1992}. An
illustration may be found in a recent study of lamellar stacks of
lipid bilayers~\cite{Bougis2015,LeiteRubim2016,Bougis2016}, where the
delicate interplay between ``unbinding'' tendencies and interactions
favouring ``bound'' systems was varied by controlling the bilayer
molecular composition. This amounted to varying simultaneously the
bending modulus $\kappa$ (``unbinding'' tendencies) and the virial
coefficient $\chi$ that encapsulates in the model the effect of
interactions~\cite{Milner1992,Bougis2015,LeiteRubim2016,Bougis2016},
\added{in a way somehow similar to the theoretical approach to the
  lamellar--lamellar phase coexistence proposed in
  ref.~}\cite{Noro1999}.
\par
\begin{figure}[htb]
  \centering
  \includegraphics[width=0.45\textwidth]{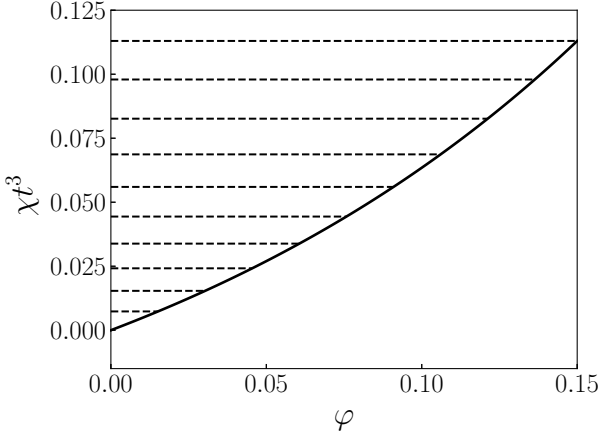}
  \caption{Phase diagram of a stack of GO sheets in the ($\varphi$,
    $\chi$) plane drawn in the case where bending modulus
    $\kappa/(k_BT)=1.$ The \textsc{Milner--Roux} virial coefficient
    $\chi$ is made dimensionless by normalising to the volume built
    from the sheet thickness $t$. Horizontal dashed lines are
    binodals, linking excess solvent with a ``bound'' lamellar stack}
  \label{fig:chiPhi}
\end{figure}

 As regards the salt effect on GO stacks, interpretations may be
simpler \added{than in refs.~}\cite{LeiteRubim2016,Bougis2016}, at
least if it is safe to assume that the main effect of salt (screening
electrostatic repulsive interactions \added{through the decrease of
  the Debye screening length}) falls upon the parameter $\chi$
\emph{only}, $\kappa$ being therefore unaffected. In such a simple
limit, the \textsc{Milner--Roux} virial coefficient $\chi$ should be a
monotonously increasing function of $c_s$, since \textsc{van der
  Waals} attractions between GO sheets would be less and less
counterbalanced by electrostatic repulsions~\cite{Milner1992},
\added{as classically described for colloidal particles in the DLVO
  theory~}\cite{ninham1999}. The thermodynamic analysis of the
unbinding transition then leads to a (schematic) phase diagram,
displayed in the ($\varphi$, $\chi$)-plane in fig.~\ref{fig:chiPhi},
\added{following refs.~}\cite{Milner1992,Nallet2017}.
\par The general features of the phase diagram are in qualitative
agreement with available observations. As long as the salt
concentration $c_s$ is low enough, interactions between GO sheets are
essentially repulsive, $\chi$ should remain ``small'' (possibly
negative) and the system is homogeneous--blank region in
fig.~\ref{fig:chiPhi}. \added{In this case}, for any given $\varphi$,
$\ell$ cannot depend on $c_s$ and obtains according to
eq.~(\ref{eq:1Ddil}) as $\ell=t/\varphi$. However, when $c_s$
increases above a threshold concentration $c_s^*(\varphi)$,
\textsc{van der Waals} attractions start being dominant in the sense
that the virial coefficient $\chi(c_s)$ \added{becomes larger than}
the swelling limit line \added{$\chi(\varphi)$ drawn in
  fig.~\ref{fig:chiPhi}}. For the same given overall composition
$\varphi$, the swollen stack of GO sheets phase-separates, part of the
volume being filled with \emph{pure} solvent \added{($\varphi_l=0$)},
a \emph{more concentrated} GO--solvent system \added{with
  $\varphi_r\geq\varphi$} occupying the remaining volume--left- and
right-end of dashed binodals in fig.~\ref{fig:chiPhi}. Since $\ell$
remains equal to $t/\varphi_r$ in the swollen stack, the observed
stacking period starts \emph{decreasing}. \added{Because the swelling
  limit line $\chi(\varphi)$ in fig.~\ref{fig:chiPhi} increases with
  $\varphi$,} the phase separation phenomenon occurs earlier
(\emph{i.e.}  for a lower salt content) if the lamellar stack is
initially more dilute.
\par To proceed further in quantitative terms, it would be desirable
to directly measure the properties controlling the swelling behaviour
in GO stacks, \emph{viz.} the bending modulus $\kappa$ of the GO sheet
and the sheet--sheet interaction potential or, at least, the
\textsc{Milner--Roux} virial coefficient $\chi$~\cite{Milner1992},
\added{in particular as a function of $c_s$}. As an intermediate step
before reaching this ultimate goal, we propose below a method (based
upon a model description of the small-angle x-ray--or
neutron--diffractograms) for estimating the \textsc{Caillé} exponent
$\eta$. This parameter was originally introduced for describing
elastic fluctuations in smectic~A liquid
crystals~\cite{Caille1972,Gunther1980,Gennes1993} and is related to
both smectic layer flexibility and interactions. It also proved useful
in interpreting characteristic features of diffractograms of lyotropic
lamellar L$_{\alpha}$ phases see, \emph{e.g.},
\cite{Safinya1986,artNallet93,Zhang1994,Bougis2015}, \added{as well as
  of GO stacks}~\cite{XuGao2011_2}.
\par The intensity $I$ of the radiation scattered by unoriented
(``powder'') lamellar samples can be shown to a good approximation to
be proportional to the product of two main
terms~\cite{bookKratkyPorod1949,artNallet93}
\begin{equation}
\label{eq:intensity}
I\left( q\right) = A\frac{2\pi}{q ^2}P\left( q\right)S\left( q\right)
\end{equation}
where $P$ and $S$ are, respectively, the form and structure factors,
accounting for the scattering along their normal by isolated flat
``particles'' and, along the stacking axis, by a 1D periodic
structure. In eq.~(\ref{eq:intensity}), $q$ is the magnitude of the
scattering wave vector and $A$ is a normalising constant that depends
on ``particle''--solvent contrast, composition, etc. The $1/q^2$ term
accounts at large enough wave vectors for the powder
average~\cite{bookKratkyPorod1949}, \added{and can also be considered
  as the ``particle'' \emph{in-plane} form factor~}\cite{artNallet93}.
\par With the further simplification of considering the GO sheets as
\emph{zero-thickness} ``particles'', the form factor $P$ no longer
depends on $q$ and remains equal to 1, \added{an acceptable
  approximation in the investigated SAXS range}.
On the other hand, the structure factor is conveniently expressed as
results from the following equations~(\ref{eq:structure}) and
(\ref{eq:alpha})
\begin{multline}
\label{eq:structure}
S\left(q \right) = 1 + 2 \sum\limits_{1}^{N-1} \left(1 -
  \frac{n}{N}\right) \\ \times\cos\left[\frac{q\ell n}{1 + 2
    \Delta q^2\ell^2\alpha\left(n\right)}\right] \\
\times\frac{\exp\left\{-\frac{2q^2\ell^2\alpha\left(n\right) + \Delta
      q^2\ell^2n^2}{2\left[1 + 2\Delta q^2\ell^2\alpha
        \left(n\right)\right]}\right\}}{\sqrt{1 + 2\Delta q^2
    \ell^2\alpha\left(n\right)}}
\end{multline}
\begin{equation}
\label{eq:alpha}
\alpha\left(n\right) = \frac{\eta}{4\pi^2}\left[\log\left(\pi n\right)
  + \gamma \right]
\end{equation}
where $N$ is the number of correlated GO sheets in the lamellar stack,
$\ell$ the period of the structure, $\Delta q$ the Gaussian width of
the resolution (or FWHM$/\sqrt{8\ln2}$), and
$\gamma\approx0.57721\ldots$ the value of \textsc{Euler}'s
constant~\cite{artNallet93}. Note that owing to the logarithmic term
in eq.~(\ref{eq:alpha}), characteristic of the anomalous fluctuation
properties in one-dimensional
systems~\cite{Caille1972,Gunther1980,Gennes1993}, the structure factor
given in eq.~(\ref{eq:structure}) differs essentially from the results
relevant for the so-called disorders of the first or second kinds, or
para-crystalline theory--see, \emph{e.g.},
\cite{Shibayama1986,Zhang1994}.
\par
\begin{figure}[h]
  \begin{center}
    \includegraphics[width=0.45\textwidth]{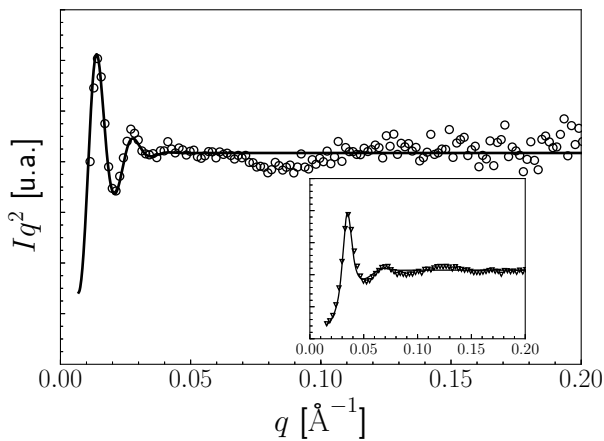}
    \caption{SAXS spectra in the Kratky representation for a
      \added{$f_m=1.4$~\%} GO dispersion in pure water ($\circ$). The
      full line is the best fit of eq.~(\ref{eq:structure}) to the
      data. Inset: \added{$f_m=4.3$~\%} system ($\triangledown$)}
    \label{fig:fitting}
  \end{center}
\end{figure}

 The model, though being somehow equivocal because the resolution
of our experiment is limited, the distinction between (small) $N$ and
(large) $\eta$ roles therefore becoming less clear-cut in some cases,
has nevertheless been used to tentatively describe the diffractograms
for \added{some rather dilute samples (without adding salt), with GO
  mass fractions $f_m=1.4$~\%, $2.8$~\%, $4.3$~\% and $7.1$~\% (volume
  fractions respectively $\varphi\approx0.8$~\%, $\approx1.6$~\%,
  $\approx2.4$~\% and $\approx4.1$~\%)}. Figure~\ref{fig:fitting}
displays \added{two} results, \added{and} fitting parameters are given
in table~\ref{tab:fit}.
\begin{table}[htb]
  \centering
  \caption{\added{Model parameters}}
  \begin{tabular}{*{5}{| >{\scriptsize}c} |}
    \hline
    Parameter & $f_m=0.014$ & $f_m=0.028$ & $f_m=0.043$ & $f_m=0.071$ \\ \hline 
    $\ell$ [nm] & 42.7 & 23.8 & 17.4 & 10.2 \\ \hline
    $\eta$ & 0.39 & 0.75 & 0.65 & 0.93 \\ \hline
    $N$ & 20 & 10 & 10 & 7 \\
  \hline
  \end{tabular}
  \label{tab:fit}
\end{table}
\par A fair description of the small-angle scattering features is
obtained when using the proposed model, with nevertheless obvious
shortcomings for scattering wave vectors in the range
$\approx0.08-0.16$~\AA$^{-1}$ that may result from the crudeness of
our assumption as regards the GO sheet form factor. In particular, the
dangling oxygen-rich groups present in GO
sheets 
\added{may increase locally the sheet thickness, therefore
  contributing to out-of-plane features of the form factor not
  accounted for in our simplified description.}
\par From the fitted values of the \textsc{Caillé} exponent $\eta$, the smectic
compression modulus $B$ of the lamellar structure made of stacked GO
sheets may be estimated. With
\begin{equation}
  \label{eq:caille}
  \eta=\frac{q_0^2k_BT}{8\pi\sqrt{KB}}
\end{equation}
\added{from ref.~}\cite{Caille1972}, and using for the smectic splay
modulus $K$ the relation $K=\kappa/\ell$~\cite{Helfrich1978}, we get
\begin{equation}
  \label{eq:Bmodulus}
  \frac{\ell^3B}{k_BT}=\frac{\pi^2k_BT}{4\kappa\eta^2}
\end{equation}
or \added{$B\approx6$ for $f_m=4.3$~\% (respectively, $B\approx16$ for
  $f_m=1.4$~\%)} in $k_BT/\ell^3$ units if, as proposed in
ref.~\cite{Poulin2016}, the value of the GO sheet bending modulus
$\kappa$ is actually equal to $k_BT$. Such values for the smectic
compression modulus $B$, significantly \emph{larger} than predicted in
the \textsc{Helfrich} model, namely
$\ell^3B_H/(k_BT)=9\pi^2k_BT/(64\kappa)$~\cite{Helfrich1978,Safinya1986}
(or $\approx1.4$ in $k_BT/\ell^3$ units), are quite reasonable in the
presence of dominantly \emph{repulsive} interactions between GO
sheets. Indeed, from \textsc{Milner-Roux} analysis of the
``unbinding'' transition~\cite{Milner1992}, the smectic compression
modulus $B$ should be expressed as~\cite{Bougis2015}
\begin{equation}
  \label{eq:unbindingB}
  \frac{\ell^3B}{k_BT}=\frac{9\pi^2k_BT}{64\kappa(1-t/\ell)^4}-2\chi\ell t^2
\end{equation}
which, from eq.~(\ref{eq:Bmodulus}) with $\ell$ and $\eta$ values as
given in table~\ref{tab:fit}, yields roughly the same estimate for the
virial coefficient \added{$\chi t^3\approx-0.04$} for the two GO
concentrations, with a \emph{negative} sign as expected for overall
repulsive interactions.
\par
\begin{figure}[h]
  \begin{center}
    \includegraphics[width=0.45\textwidth]{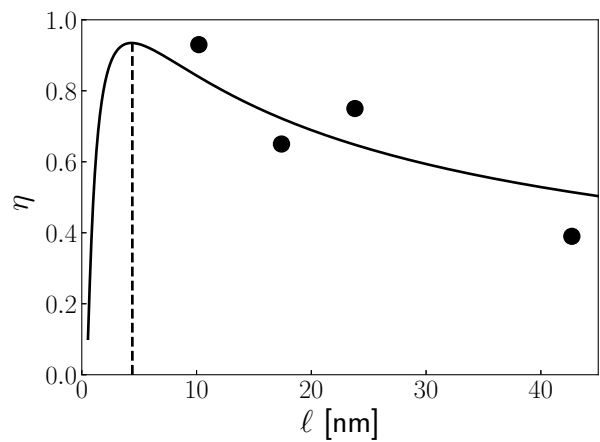}
    \caption{\textsc{Caillé} exponent along \added{the dilution line
        with pure water}, as resulting from
      equations~(\ref{eq:caille}) and (\ref{eq:unbindingB}), with
      $\kappa/k_BT=1$, \added{$\chi t^3=-0.036$}. GO sheet thickness
      \added{$t=0.39$~nm}. Vertical dashed line drawn at
      \added{$\ell^*=5.4$~nm}. Data points from table~\ref{tab:fit}}
    \label{fig:etaMilnerRoux}
  \end{center}
\end{figure}

 The structural phase transition that occurs in the vicinity of
$\ell=2.5$~nm is actually also amenable, qualitatively at least, to an
interpretation in terms of \textsc{Milner-Roux} arguments. As shown in
fig.~\ref{fig:etaMilnerRoux},
the \textsc{Caillé} exponent is expected to strongly \emph{increase}
when the lamellar stack of GO sheets is dehydrated, until a
characteristic period $\ell^*$ is reached beyond which $\eta$
decreases to very small values. As larger $\eta$ values are associated
to (lamellar) Bragg peaks with lesser peak intensities and broader
tails, the quasi-disappearance of the first and second order Bragg
peaks in a given dilution range (see fig.~\ref{fig:porodTransition})
may thus be understood, even though the predicted $\ell^*$ value,
namely \added{5.4~nm}, clearly differs from its experimental
counterpart. However, since the \textsc{Milner-Roux} description of
the unbinding transition is a mean-field, \emph{perturbative} theory,
we believe that such a discrepancy should not be too seriously
deplored for such concentration ranges where direct interactions
between GO sheets are definitely strong.
\section{Conclusion}
\label{conclusion}
A procedure to concentrate aqueous GO dispersions to significant
dryness, with the benefit of avoiding the formation of aggregates has
been implemented. The lamellar stacks of GO sheets obtained in an
extended concentration range, from \emph{ca.} 2~\% to \added{72~\%},
can be reversibly swollen or dehydrated. The simple one-dimensional
dilution law is largely obeyed over all the investigated hydration
range, even though conspicuous discrepancies have been revealed by
small-angle x-ray scattering studies that may indicate the occurrence
of an underlying, as yet unidentified, structural phase transition.
\par The swelling behaviour of the aqueous GO dispersions can be
interpreted, similarly to many lyotropic lamellar L$_{\alpha}$ phases
in amphiphilic systems, in terms of an entropic ``force'' arising from
the confinement of undulation fluctuations (also known as
\textsc{Helfrich} undulation interactions) acting together, or
competing with, direct forces, respectively electrostatic repulsions
and \textsc{van der Waals} attractions. The so-called ``unbinding
transition'' mechanism appears here to be mainly driven by the ionic
strength of the aqueous medium swelling the GO sheets, as
\added{indirectly} suggested by the quantitative analysis of the
small-angle x-ray diffractograms in terms of a parameter, the
\textsc{Caillé} exponent $\eta$, that combines the bending and
compression moduli characterising the elastic properties of lamellar
phases. The analysis confirms the recently proposed ``super-flexible''
nature of GO sheets~\cite{Poulin2016}.
%
%
\begin{acknowledgement}
  We thank Philippe Poulin, C\'ecile Zakri and Wilfrid Neri for
  fruitful discussions. We also thank Conselho Nacional de
  Desenvolvimento Cient\'ifico e Tecnol\'ogico (CNPq - Brazil) for
  providing support to Rafael Leite Rubim through the program
  ``Ci\^encia sem Fronteiras'' (processo CNPq
  n\textordmasculine~250085/2013-5).
\par Figures have been drawn with the 2D graphics package
\texttt{Matplotlib}~\cite{Hunter2007}
\end{acknowledgement}
\section{Author contribution statement}
\added{L.N., and F.N. conceived and planned the
  research. R.L.R. conceived and implemented the drying
  procedure. R.L.R., M.A.B., T.M., and K.B. contributed to sample
  preparation and carried out the experiments. All authors provided
  critical feedback and helped shape the research, analysis and
  manuscript. All the authors have read and approved the final
  manuscript.}
\bibliographystyle{unsrt}
\bibliography{bibliographieGO}
\end{document}